\documentstyle[sprocl,epsfig]{article}

\def\ra{\rightarrow}

\def\be{\begin{equation}}
\def\ee{\end{equation}}
\def\bea{\begin{eqnarray}}
\def\eea{\end{eqnarray}}
\def\cent{\centerline}

\def\vs{\vskip}
\def\\{\hfill\break}

\def\pa{\parindent}

\def\ej{\vfill\eject}
\def\ran{\rangle}
\def\lan{\langle}
\def\hf{\hfill}

\newcommand{\kos}{\ifmmode \mathrm{K^{0}_{S}} \else K$^{0}_{\mathrm S} $ \fi}
\newcommand{\kol}{\ifmmode \mathrm{K^{0}_{L}} \else K$^{0}_{\mathrm L} $ \fi}
\newcommand{\kob}{\ifmmode {\overline{\mathrm{K}^{0}}} 
\else $\overline{\mathrm{K}^{0}} $ \fi}

\def\ifmath#1{\relax\ifmmode #1\else $#1$\fi}%
\def\ra{{\mathrm{a}}}
\def\rb{{\mathrm{b}}}
\def\rc{{\mathrm{c}}}
\def\rd{{\mathrm{d}}}
\def\re{{\mathrm{e}}}
\def\rf{{\mathrm{f}}}
\def\rg{{\mathrm{g}}}
\def\rh{{\mathrm{h}}}
\def\rd{\ifmath{{\mathrm{d}}}}
\def\rG{\ifmath{{\mathrm{G}}}}
\def\rK{\ifmath{{\mathrm{K}}}}
\def\rL{\ifmath{{\mathrm{L}}}}
\def\ro{\ifmath{{\mathrm{o}}}}
\def\rp{\ifmath{{\mathrm{p}}}}
\def\rt{\ifmath{{\mathrm{t}}}}
\def\rT{\ifmath{{\mathrm{T}}}}
\def\rs{\ifmath{{\mathrm{s}}}}

\def\ev{\ifmath{{\mathrm{ev}}}}
\def\eff{\ifmath{{\mathrm{eff}}}}
\def\rms{\ifmath{{\mathrm{rms}}}}
\def\side{\ifmath{{\mathrm{side}}}}
\def\out{\ifmath{{\mathrm{out}}}}
\def\vec#1{{\mbox{\bf #1}}}

\newcommand{\beq}{\begin{equation}}
\newcommand{\eeq}{\end{equation}  }
\newcommand{\beqa}{\begin{eqnarray}}
\newcommand{\eeqa}{\end{eqnarray}  }

\begin{document}

\hf Nijmegen preprint

\hf HEN-405

\hf Dec. 97

\vs 1cm
{\pa=0pt
{\bf ESTIMATION OF HYDRODYNAMICAL MODEL PARAMETERS FROM\\ 
THE INVARIANT SPECTRUM AND THE BOSE-EINSTEIN CORRELATIONS OF\\ 
$\pi^-$ MESONS PRODUCED IN $(\pi^+/\rK^+)$p INTERACTIONS AT 250 GeV/$c$}
\vs 5mm
\cent{EHS/NA22 Collaboration}
\vs 5mm
N.M. Agababyan$^{\rg}$, M.R. Atayan$^{\rg}$, T. Cs\"org\H{o}$^{\rh}$,
E.A. De Wolf$^{\ra,1}$,
K. Dziunikowska$^{\rb,2}$, A.M.F. Endler$^{\re}$,\\
Z.Sh. Garutchava$^{\rf}$,
H.R. Gulkanyan$^{\rg}$, R.Sh. Hakobyan$^{\rg}$, J.K. Karamyan$^{\rg}$, 
D. Kisielewska$^{\rb,2}$,\\
 W. Kittel$^{\rd}$, S.S. Mehrabyan$^{\rg}$, Z.V. Metreveli$^{\rf}$, 
K. Olkiewicz$^{\rb,2}$, F.K. Rizatdinova$^{\rc}$,\\ 
E.K. Shabalina$^{\rc}$, L.N. Smirnova$^{\rc}$, M.D. Tabidze$^{\rf}$, 
L.A. Tikhonova$^{\rc}$, A.V. Tkabladze$^{\rf}$,\\ 
A.G. Tomaradze$^{\rf}$, F. Verbeure$^{\ra}$, S.A. Zotkin$^{\rc}$
 
\begin{itemize}
\itemsep=-2mm
\item[$^\ra$] Department of Physics, Universitaire Instelling Antwerpen, 
B-2610 Wilrijk, Belgium
\item[$^\rb$] Institute of Physics and Nuclear Techniques of Academy of
Mining and Metallurgy and Institute of Nuclear Physics, PL-30055 Krakow,
Poland
\item[$^\rc$] Nuclear Physics Institute, Moscow State University, RU-119899
Moscow, Russia
\item[$^\rd$] High Energy Physics Institute Nijmegen (HEFIN),
University of Nijmegen/NIKHEF, NL-6525 ED Nijmegen, The
Netherlands
\item[$^\re$] Centro Brasileiro de Pesquisas Fisicas, BR-22290 Rio de Janeiro,
Brazil
\item[$^\rf$] Institute for High Energy Physics of Tbilisi State University,
GE-380086 Tbilisi, Georgia
\item[$^\rg$] Institute of Physics, AM-375036 Yerevan, Armenia
\item[$^\rh$] KFKI, Hungarian Academy of Sciences, H-1525 Budapest 114,
Hungary
\end{itemize}

\vs 1cm
{\bf Abstract:}
 The invariant spectra of $\pi^-$ mesons produced in 
$(\pi^+/\rK^+)\rp$ interactions at 250 GeV/$c$ are
analysed in the framework of the hydrodynamical model of
three-dimensionally expanding cylindrically symmetric finite systems. 
 A satisfactory description of  experimental data is achieved.
 The data favour the pattern according to which the hadron matter 
undergoes predominantly longitudinal expansion and non-relativistic
transverse expansion with mean transverse velocity $\langle u_\rt \rangle = 
0.20\pm0.07$, and is characterized by a large temperature
inhomogeneity in the transverse direction: the extracted freeze-out 
temperature at the center of the tube and at the transverse rms radius
are $140\pm3$ MeV and $82\pm7$ MeV, respectively.  The width of the 
(longitudinal) space-time rapidity distribution of the pion 
source is found to be $\Delta \eta = 1.36 \pm 0.02$. 
 Combining this estimate with results of the Bose-Einstein correlation 
analysis in the same experiment, one extracts a mean freeze-out 
time of the source of $\langle \tau_\rf \rangle = 1.4\pm0.1$ fm/$c$ and its 
transverse geometrical rms radius, $R_\rG(\rms)=1.2\pm0.2$ fm.

\vfill
\hrule width3truecm
\vs 2mm
$^1$ Onderzoeksdirecteur NFWO, Belgium\\
$^2$ Supported by the Polish State Committee for Scientific Research
\par}

\newpage

\section{Introduction}

Recent investigations of interference correlations of identical mesons 
(Bose-Einstein correlations, BEC) give evidence for a collective nature of 
the evolution of hadronic matter created in high-energy heavy ion collisions 
[1--4]. The hadronic matter flow (developed predominantly in the 
longitudinal direction) gives rise to dynamical correlations between 
space-time and phase-space coordinates of produced particles. As a 
consequence, the 'interferometric' radius, determining the shape of the 
two-particle correlation function at small relative momenta, turns out 
to be dependent on the kinematical variables of the particle pair.
It does not measure the genuine geometrical size of the meson-emitting 
source, but the effective size of the source segment radiating mesons 
with sufficiently small relative momentum. The effective size reflects 
both the geometrical size and the 'homogeneity' length of the source. 
The latter is determined by macroscopic properties as temperature and 
expanding velocity profiles of the hadronic matter at freeze-out or 
hadronization time.

These expectations are based on hydrodynamical models 
(see [5--7] and references therein) and are qualitatively confirmed by 
experimental data concerning interferometric radii and their dependence 
on the di-meson rapidity and transverse mass. It is interesting, that 
this observation not only holds for heavy-ion collisions [1--4], 
but also for meson-proton collisions [8,9] for which the formation of 
hadronic matter with properties of macroscopic (hydrodynamical) 
systems is less evident. 

One should stress, however, that correlation measurements do not contain
the complete information on the geometrical and dynamical parameters 
characterizing the evolution of the hadronic matter. More comprehensive 
information can be provided by a combined analysis of data on 
two-particle correlations and single-particle inclusive spectra [5--7].

This work is, therefore,  devoted to an analysis, in the framework of the 
hydrodynamical model for three-dimensionally expanding systems [7], of 
the invariant spectra of $\pi^-$ mesons in the central rapidity region of    
$(\pi^+/\rK^+)\rp$ interactions at $250$ GeV/$c$ and its combination with 
earlier results [8,9] on two-particle correlations. The measurements are 
performed with the help of the European Hybrid Spectrometer (experiment 
NA22) at the CERN SPS. The first NA22 data on the pion inclusive spectra 
[10] (with about half the statistics used in the present paper) show their 
similarity in $\pi^+$ and $\rK^+$ induced reactions. The present analysis,
therefore, is based on the combined $\pi^+\rp$ and $\rK^+\rp$ data.
Earlier results on BEC in the same data can be found in [8,9,11-14].

The hydrodynamical model parametrizations are discussed in Sect.2. 
Experimental data and the results of their analysis are presented 
and combined with the results from two-particle correlations in Sect.3. 
Conclusions are summarized in Sect.4.

\section{The parametrizations of invariant pion spectra}

The experimental data are analyzed in the framework of the hydrodynamical 
model for three-dimensio\-nally expanding cylindrically symmetric systems [7].
There, the invariant spectrum of pions in rapidity $y$ and transverse mass 
$m_\rt$ is approximated by 

\begin{eqnarray}
f(y,m_\rt)&=&\frac{1}{N_{\ev}}\frac{\rd N_{\pi}}{\rd y \rd m_\rt^2} = 
\nonumber \\
  &=& C {m_\rt}^{\alpha}\cosh\eta_s\exp\left(\frac{\Delta 
\eta_*^2}{2}\right)\exp\left[-\frac{(y-y_0)^2}{2\Delta y^2}\right]
\exp\left(-\frac{m_\rt}{T_0}\right) \times \nonumber  \\ 
 && \times \exp\left\{\frac{\langle 
u_\rt\rangle^2(m_\rt^2-m_{\pi}^2)}{2T_0[T_0+(\langle u_\rt\rangle^2+\langle
\frac{\Delta T}{T}\rangle)m_\rt]}\right\}.  
\end{eqnarray}
with
\be
\Delta y^2 = \Delta \eta^2 + \frac{T_0}{m_\rt} \\
\ee
\be
\frac{1}{\Delta \eta_*^2} = \frac{1}{\Delta \eta^2} + 
\frac{m_\rt}{T_0}\cosh\eta_\rs, \\
\ee
\be
\eta_\rs = \frac{y-y_0}{1+\Delta \eta^2 \frac{m_\rt}{T_0}}. \\ 
\ee

The width $\Delta y$ of the rapidity distribution given by (2) is determined 
by the width $\Delta \eta$ of the longitudinal space-time rapidity $\eta$ 
distribution of the pion emitters and by the thermal smearing width 
$\sqrt{T_0/m_\rt}$, where $T_0$ is the freeze-out temperature (at 
the mean freeze-out time $\tau_\rf$) at the axis of the hydrodynamical 
tube, $T_0=T_\rf(r_\rT=0)$. For the case of a slowly expanding system one 
expects $\Delta \eta \ll T_0/m_\rt$, while for the case of a
relativistic longitudinal expansion the geometrical extension $\Delta 
\eta$ can be much larger than the thermal smearing (provided $m_\rt>T_0$).  

Except for the inhomogeneity caused by the longitudinal expansion, (1) also 
considers the inhomogeneity related to the transverse
expansion (with the mean radial component $\langle u_\rt\rangle$ of 
hydrodynamical four-velocity) and to the transverse temperature 
inhomogeneity, characterized by the quantity
\be
\biggl\lan \frac{\Delta T}{T} \biggr\ran = \frac{T_0}{T_{\rms}}-1,
\ee
where $T_{\rms}=T_\rf(r_\rT=r_\rT(\rms))$ is the freeze-out temperature at the 
transverse rms radius $r_\rT(\rms)$ and at time $\tau_\rf$.

The exponential parameter $\alpha$ in (1) is related [7] to the number 
$k$ of dimensions in which the expanding system is inhomogeneous. For the 
special case of the one-dimensional inhomogeneity ($k=1$) caused by the 
longitudinal expansion, $\alpha=1-0.5k=0.5$ (provided $\Delta 
\eta^2\gg{T_0}/{m_\rt})$. The transverse inhomogeneity of the system 
leads to smaller values of $\alpha$. The minimum value of $\alpha=-1$ is 
achieved at $k=4$ for the special case of a three-dimensionally 
expanding system with temporal change of local temperature during the 
particle emission process.

The parameter $y_0$ in (1) denotes the midrapidity in the interaction 
c.m.s. and can slightly differ from $0$ due to different species of 
colliding particles. The parameter $C$ is an overall normalization 
coefficient.

Note that (1) yields the single-particle spectra of the
core (the central part of the interaction that supposedly
undergoes collective expansion). However, long-lived resonances
also contribute to the single-particle spectra through their
decay-products. Their contribution can be determined in the 
core-halo picture [20] by the momentum dependence of the
intercept parameter $\lambda(y,m_\rt)$ of the two-particle
Bose-Einstein correlation function. We determined
this quantity in [8] in two different $m_\rt$ windows for the reaction
investigated in the present paper, and found that within 
three standard deviations the $\lambda (m_\rt)$ parameter
is approximately independent of $m_\rt$. Hence this correction
can be absorbed in the overall normalization. 

The two-dimensional distribution (1) can be simplified for 
one-dimensional slices [7,15]:

1. At fixed $m_\rt$, the rapidity distribution reduces to the approximate 
parametrization 
\be
f(y,m_\rt) = C_m \exp\left[ -\frac{(y-y_0)^2}{2\Delta y^2}\right],
\ee
where $C_m$ is an $m_\rt$-dependent normalization coefficient and $y_0$ 
is defined above. The width parameter $\Delta y^2$ extracted for 
different $m_\rt$-slices is predicted to depend linearly on 
$1/m_\rt$, with slope $T_0$ and intercept $\Delta \eta^2$ (cf. (2)).

Note, that for static fireballs or spherically expanding shells
(6) and (2) are satisfied with $\Delta\eta = 0$  [15]. Hence the
experimental determination of the $1/m_\rt$ dependence of the $\Delta y$
parameter can be utilized to distinguish between longitudinally
expanding finite systems versus static fireballs or spherically expanding
shells.

2. At fixed $y$, the $m_\rt^2$-distribution reduces to the approximate 
parametrization
\be
f(y,m_\rt) = C_y m_\rt^{\alpha} \exp\left(-\frac{m_\rt}{T_{\eff}}\right)
\ee
where $C_y$ is a $y$-dependent normalization coefficient and $\alpha$ 
is defined as above.

The $y$-dependent "effective temperature" $T_{\eff}(y)$ can be 
approximated as
\be
T_{\eff}(y) = \frac{T_*}{1+a(y-y_0)^2} ,
\ee
where $T_*$ is the maximum of $T_{\eff}(y)$ achieved at $y=y_0$, and
\be
a = \frac{T_0 T_*}{2 m_{\pi}^2(\Delta \eta^2 + \frac{T_0}{m_{\pi}})^2}
\ee
with $T_0$ and $\Delta \eta^2$ as defined above.

The approximations (6) and (7) explicitly predict a specific narrowing 
of the rapidity and transverse mass 
spectra with increasing $m_\rt$ and $y$, respectively (cf. (2) and 
(8)). The character of these variations is expected [15] to be different 
for the various scenarios of hadron matter evolution.    

\section{\bf The results}
\subsection{The data}

The experimental setup and the data-handling procedure are described in 
[10,16]. The results of this paper are based on the analysis of 104145 
events of $(\pi^+/\rK^+)\rp$ interactions containing at least one negative 
track with momentum resolution better than $4\%$ (depending on the 
momentum, the average momentum resolution varies from $1\%$ to to $2.5\%$). 
All negative particles are assumed to have the pion mass. The contamination 
from other particles is estimated to be $(7\pm3)\%$ [10].
Candidates for single diffraction dissociation with charge 
multiplicity $n\leq6$ are excluded. For each event, a weight is 
introduced in order to normalize to the non-single diffractive 
topological cross sections [16].

The analysis is restricted to centrally produced $\pi^-$-mesons with 
cms rapidity $|y|<1.5$. The total number of pions used in the analysis 
is about 397~k.

\subsection{The rapidity distribution}

The rapidity distributions fitted by (6) are given in Fig.~1 for
23 $m_\rt$ slices and $m_\rt$ values ranging from 0.16 to 0.63 GeV. 
The fit quality is satisfactory for all slices ($0.7<\chi^2/NDF<1.5$). 
The fitted $\Delta y^2$ values plotted in Fig.~2 demonstrate the widening 
of the $y$-distribution with increasing $1/m_\rt$. 
According to (2), the linear dependence of $\Delta y^2$ on 
$1/m_\rt$ is described by intercept 
$\Delta \eta^2 = 1.91\pm 0.12$ and 
slope $T_0=159\pm 38$MeV. Thus, the width of the $y$-distribution is 
dominated by the spatial (longitudinal) distribution of pion emitters 
(inherent to longitudinally expanding systems) and not by the thermal 
properties of the hadron matter, as would be expected for static or 
radially expanding sources.

\subsection{The $m_\rt$ distribution}

The $m_\rt$-distributions fitted by (7), are given in Fig.~3 for 11 
$y$-slices for $y$-values between -1.5 to +1.5. For all slices 
$\chi^2/NDF<1.4$.
The fitted values of $T_{\eff}(y)$ (Fig.~4) tend to decrease with increasing 
$|y|$ and approximately follow parametrization 
(8) with $T_*=160\pm1$ MeV, $a=0.083\pm0.007$ and $y_0=-0.065\pm0.039$.
Note, however, that the fit of $T_{\eff}(y)$ by (8) has low confidence 
level: $\chi^2/NDF=29/11$. This large value is due to an asymmetry in the
$T_\eff$ distribution with respect to $y=0$. Except for the last
point, $T_\eff$ is higher in the meson than in the proton hemisphere.

Note, furthermore, that the small value of the 
parameter $a$ is the result of the comparatively large longitudinal 
extension $\Delta \eta $ of the pion source (cf. (9)). Note also the 
consistency between the quoted value of $a$ and the value 
$a\approx 0.07$ calculated from (9) at values of the parameters $\Delta 
\eta^2$, $T_0$ and $T_*$ extracted from two different fits of 
the one-dimensional spectra (6) and (7).

The fitted values of the exponential parameter $\alpha$ in (7) are near 
zero (varying between $-0.02\pm 0.02$ and $0.14\pm 0.02$ for different 
$y$-slices). These small values of $|\alpha |$ correspond 
to a two-dimensional inhomogeneity of the expanding system 
($\alpha = 1-0.5k$). One concludes, therefore, that apart from a 
longitudinal inhomogeneity caused by the relativistic longitudinal flow, 
the hadron matter can also possess a transverse inhomogeneity (caused by 
transverse expansion or a transverse temperature gradient) or 
undergoes a temporal change of local temperature during the particle 
emission process.

\subsection{The transverse direction}

Further information on hadron-matter evolution in the transverse 
direction can be extracted from a more detailed analysis using the 
two-dimensional parametrization (1) with parameters $\langle u_\rt\rangle $ 
and $\langle \frac{\Delta T}{T}\rangle $ characterizing the strength of 
the transverse expansion and temperature inhomogeneity.
The results of the fit are presented in Table~1.  

The fitted value of the exponential parameter of $\alpha = 0.26\pm 0.02$ 
corresponds to an effective number $k_{\eff}$ of dimensions in which the 
expanding system is inhomogeneous, $k_{\eff}\approx 1.5$, 
definitely in excess of $k=1$ corresponding to a one-dimensional 
(longitudinal) inhomogeneity. This can be at least partly attributed to 
a transverse inhomogeneity of the system. The moderate value of the 
mean transverse four-velocity $\langle u_\rt\rangle=0.20\pm 0.07$ (Table~1) 
indicates that the transverse inhomogeneity is only to small 
extent caused by a (non-relativistic) transverse expansion. It is mainly 
stipulated by a rather large temperature inhomogeneity characterized by 
the fit result of $\langle \frac{\Delta T}{T}\rangle =0.71\pm 0.14$. 
Using (5), one infers that the freeze-out temperature decreases from 
$T_0=140\pm 3$ MeV at the central axis of the hydrodynamical tube to 
$T_{\rms}=82\pm 7$ MeV at a radial distance equal to the transverse 
rms radius of the tube.

\subsection{Combination with two-particle correlations}

As already mentioned in the introduction, more comprehensive information 
on geometrical and dynamical properties of the hadron matter 
evolution are expected from a combined consideration of 
two-particle correlations and single-particle inclusive spectra [5--7].

The two-particle correlation function $K_2(\vec p_1,\vec p_2)$ at small 
momentum difference $\vec q=\vec p_1-\vec p_2$ is often approximated by a 
Gaussian function [17]:
\be
K_2(\vec p_1,\vec p_2)\sim 
\exp[-R_\rL^2q_\rL^2-R_\ro^2q_{\out}^2-R_\rs^2q_{\side}^2] ,
\ee
where the three orthogonal components $q_\rL,q_{\out},q_{\side}$ of the vector 
$\vec q$ are oriented, respectively, along the collision axis, along 
and perpendicular to the pair transverse momentum; $R_\rL,R_\ro$ and $R_\rs$ 
are, respectively, the longitudinal, 'out' and 'side' effective 
dimensions of the source segment radiating the BE correlated pion 
pairs. Due to the non-static nature of the source, these effective sizes 
vary with the average transverse mass 
$M_\rt=\frac{1}{2}(m_{\rt1}+m_{\rt2})$ and the average rapidity 
$Y=\frac{1}{2}(y_1+y_2)$ of the pion pair. In the 'longitudinal c.m.s.' 
(LCMS) [18], where $Y=0$, the effective radii can be approximately 
expressed as [7,15,19]:
\be
R_\rL^2=\tau_\rf^2\Delta \eta_*^2 \\
\ee
\be
R_\ro^2=R_*^2+\beta_\rt^2\Delta \tau_*^2  \\
\ee
\be
R_\rs^2=R_*^2
\ee        
with
\be
\frac{1}{\Delta \eta_*^2}=\frac{1}{\Delta \eta^2}+\frac{M_\rt}{T_0} \\
\ee
\be
R_*^2=\frac{R_\rG^2}{1+\frac{M_\rt}{T_0}(\langle u_\rt\rangle^2+\langle 
\frac{\Delta T}{T}\rangle)}, \\
\ee
where parameters $\Delta \eta^2,T_0,\langle u_\rt\rangle$ and $\langle 
\frac{\Delta T}{T}\rangle$ are defined and estimated above from the 
invariant spectra;
$R_\rG$ is related to the transverse geometrical rms radius of the 
source as $R_\rG(\rms)=\sqrt{2} R_\rG$;
$\tau_\rf$ is the mean freeze-out (hadronization) time;
$\Delta \tau_\ast$ is related to the duration time $\Delta \tau$ of pion 
emission and to the temporal inhomogeneity of the local temperature; if the 
latter has a small strength (as one can deduce from the restricted 
inhomogeneity dimension estimated above: $k_{\eff}\approx 1.5$), an 
approximate relation $\Delta \tau \geq \Delta \tau_*$ holds;
the variable $\beta_\rt$ is the transverse velocity of the pion pair.

Relations (11)--(15), combining the results of the 
single-particle invariant spectrum and the two-particle correlation 
function, allow one to extract additional information on the parameters 
$(\tau_\rf,\Delta \tau_*,R_\rG$) characterizing the space-time evolution of 
hadron matter.

The interferometric radii $R_\rL,R_\ro,R_\rs$ were extracted recently [8,9] in 
the same experiment from the correlations of negative pions. To match with 
parametrization (10), the radii quoted in [8,9] are divided here by a factor 
$\sqrt{2}$.

The effective longitudinal radius $R_\rL$, extracted for two 
different mass ranges, $M_\rt=0.26\pm 0.05$ and $0.45\pm 0.09$ GeV/$c^2$ 
are found to be $R_\rL=0.93\pm 0.04$ and $0.70\pm 0.09$ fm, respectively. 
This dependence on $M_\rt$ matches well the predicted one. Using (11) 
and (14) with $T_0=140\pm 3$ MeV and $\Delta \eta^2 
=1.85\pm 0.04$ (Table~1), one finds that the values of $\tau_\rf$ 
extracted for the two different $M_\rt$-regions are similar to
each other: $\tau_\rf =1.44\pm 0.12$ and $1.36\pm 0.23$ fm/$c$. 
The averaged value of the mean freeze-out 
time is $\tau_\rf =1.4\pm 0.1$ fm/$c$.

Since we find that $\Delta\eta$ is significantly bigger than 0,
static fireballs or spherically expanding shells, that were found
to be able to describe our two-particle correlation data in [8], 
fail to reproduce our single-particle spectra.

The transverse-plane radii $R_\ro$ and $R_\rs$ measured in [8,9] for the 
whole $M_\rt$ range are: $R_\ro=0.91\pm 0.08$ fm and $R_\rs=0.54\pm 0.07$ fm. 
Substituting in (12) and (13), one obtains (at $\beta_\rt=0.484c$ [8]): 
$\Delta \tau_*=1.3\pm 0.3$ fm/$c$. The mean duration time of pion 
emission can be estimated as $\Delta \tau \geq \Delta \tau_* =1.3\pm 0.3$ 
fm/$c$.
A possible interpretation of $\Delta \tau \approx \tau_\rf$ might be that 
the radiation process occurs during almost all the hydrodynamical 
evolution of the hadronic matter produced in meson-proton collisions.

An estimation for the parameter $R_\rG$ can be obtained from (13) and 
(15) using the quoted values of $R_\rs,T_0,\langle u_\rt\rangle $ and $\langle 
\frac{\Delta T}{T} \rangle $ at the mean value of $\langle M_\rt \rangle 
=0.31\pm 0.04$ GeV/$c$ (averaged over the whole $M_\rt$-range): 
$R_\rG=0.88\pm 0.13$ fm. The geometrical rms transverse radius of the 
hydrodynamical tube, $R_\rG(\rms)=\sqrt{2}R_\rG=1.2\pm 0.2$ fm, turns out 
to be larger than the proton rms transverse radius.

\subsection{Estimation of systematic errors}
The analytic formulae used to estimate
the hydrodynamical model parameters were evaluated under
certain approximations. 
As shown in [21], an accuracy better than 10-20\% is reached
under the following conditions:

a) 
\vs-5mm
\be
\frac{\beta_\rt \lan u_\rt\ran R_*^2}{\tau_0 R_\rG T_0/M_\rt} < 0.6 \ .
\ee
With our parameter values, the l.h.s. of (16) turns out
to be less than 0.13 for the whole range of $M_\rt$;

b) Additionaly, the validity of the formulae (1)-(9) for the 
invariant spectra requires:

\be
\frac{|y-y_0|}{1+\Delta\eta^2 m_\rt/T_0} < 1 \ .
\ee

With our parameter values, the l.h.s. of (17) turns out
to be less than 0.52 for the full ranges of $y$ and $m_\rt$ considered.

c) Formulae (10)-(15) for the two-particle correlation function
are derived under the additional, more stringent condition:

\be
\frac{|Y-y_0|}{1+\Delta\eta^2 (M_\rt/T_0-1)} <1 \ .
\ee
 
At the maximum value of $|Y-y_0|$, (18) is fulfilled,
except for a fraction of 8\% of pion pairs with $M_\rt<0.18$ GeV/$c^2$.
In the full range of $|Y-y_0|<1.5$ considered, this is reduced to less 
than 4\% of
pion pairs.
Note further that the upper limit of 10-20\% relative errors is reached 
only at the upper limit of
inequalities (16-18) and that the
squared relative errors increase quadratically with the increase of the
left-hand-sides of (16-l8).

Hence we estimate an upper limit of order
$10-20 \% \times (0.13/0.6) \times (0.52/1.0) \times 1.0 = 1-2\%$
for the precision of the analytic formulae used in the fits,
except for 4\% of the pion pairs where the 
relative systematic errors may reach 10-20\%.
This accuracy is well within the mean experimental (statistical) errors of 
the measured two-dimensional invariant spectra, analyzed in the present work
(about 10\%) and of the three-dimensional correlation function (about 
15\%), analyzed in [8,9].

\subsection{The space-time distribution of $\pi$ emission}
On Figure 5 we show a reconstruction of the space-time 
distribution of pion emission points, expressed as a function 
of the cms time variable $t$ and the cms longitudinal 
coordinate $z$.
The momentum-integrated emission function  along the $z$-axis,
i.e., at ${\vec r}_\rt = (r_x, r_y) = (0,0)$ is given by
\beq
 S(t,z) \propto \exp\left(-  {(\tau - \tau_0)^2\over 2 \Delta \tau^2} 
\right) \exp\left( - {(\eta - y_0)^2   \over 2 \Delta \eta^2} \right).
\eeq
It relates the parameters fitted to our data with particle production 
in space-time. Note that the coordinates $(t,z)$, 
can be expressed with the help of the longitudinal proper-time $\tau$
and space-time rapidity $\eta$ as $(\tau \cosh(\eta), \tau \sinh(\eta) )$.

We find a structure looking 
like a boomerang, i.e., particle production takes place close to the
regions of $z=t$ and $z=-t$, with gradually decreasing
probability for ever larger values of space-time 
rapidity. Although the mean proper-time for particle
production is $\tau_\rf=1.4$ fm/$c$, and the dispersion
of particle production in space-time rapidity is
rather small, $\Delta \eta = 1.35$ fm, we still see a characteristic
long tail of particle emission on both sides of the 
light-cone, giving a total of 40 fm maximal longitudinal
extension in $z$  and a maximum of about 20 fm/$c$
duration of particle production in the time variable $t$.

\section{Summary} 

The invariant spectra of $\pi^-$-mesons produced in 
$(\pi^+/\rK^+)\rp$-interactions at 250 GeV/$c$ are analyzed in the framework 
of the hydrodynamical model of a three-dimensionally expanding 
cylindrically symmetric finite system [7].

The data favour a picture according to which the hadron matter 
undergoes extensive longitudinal expansion with a space-time rapidity 
width of $\Delta \eta =1.36\pm 0.02$ and a non-relativistic transverse 
expansion with the mean transverse four-velocity $\langle u_\rt \rangle 
=0.20\pm 0.07$.

The hadron matter possesses a large temperature inhomogeneity in the 
transverse direction: the freeze-out temperature varies from $T_0=140\pm 
3$ MeV at the central axis of the hydrodynamical tube to $T_{\rms}=82\pm 
7$ MeV at a radial distance equal to the transverse rms radius of 
the tube.

The information from the single-particle invariant spectrum is combined
with that from the two-particle correlation function.
The transverse mass dependence of the width $\Delta y$ 
of the single-particle $y$-distribution and that of the interferometric 
longitudinal radius $R_\rL$ are 
found to be consistent. This allows to extract the mean freeze-out  
(hadronization) time of the strongly interacting matter: $\tau_\rf=1.4\pm 
0.1$ fm/$c$. The duration time of pion emission is estimated to be 
$\Delta \tau \geq 1.3\pm 0.3$ fm/$c$, i.e. close to $\tau_\rf$, indicating 
that the emission process occurs during almost all the hydrodynamical 
evolution. As an estimate for the transverse geometrical rms radius of the 
hydrodynamical tube we obtain : $ R_\rG(\rms)=1.2\pm 0.2$ fm.

The combined analysis of single-particle spectra and Bose-Einstein 
correlation functions increases the selective power of the analysis. 
Static (Kopylov-Podgoretskii) fireballs and spherically expanding shells, 
that were found to be able to reproduce our interferometric data in [8] 
fail to reproduce simultaneously the single-particle spectra and the BE 
correlations in the NA22 experiment. 

The systematic errors on the quoted parameters related to the limited 
accuracy of the approximate analytic formulae (used in the present
study) are estimated to be of the order of or smaller than the
quoted statistical errors.

As far as we know, this is the first time that a full reconstruction of 
the space-time distribution of the particle emitting source is obtained 
in hadron-hadron reactions from a combined analysis of 
single-particle spectra and Bose-Einstein correlation measurements. 

\subsection*{Acknowledgments}
We are grateful to J.D. Bjorken for encouragement on this topic
and to B. L\"orstad for suggestions.
We are grateful to the III. Physikalisches Institut B, RWTH
Aachen, Germany, the DESY-Institut f\"ur Hochenergiephysik, Berlin-Zeuthen,
Germany, the Institute for High Energy Physics, Protvino, Russia,
the Department of High Energy Physics, Helsinki University, Finland,
and the University of Warsaw and Institute of Nuclear Problems, Poland for
early contributions to this experiment. 
This work is part of the research program of the ``Stichting
voor Fundamenteel Onderzoek der Materie (FOM)", which is financially
supported by the ``Nederlandse Organisatie voor Wetenschappelijk Onderzoek
(NWO)". We further thank NWO for support of this project within the program
for subsistence to the former Soviet Union (07-13-038). 
The activity of the Yerphi group is partially supported by the Government 
of Republic of Armenia, in the framework of the theme no.94-496. 

This work was partially supported by the Hungarian NSF grants OTKA -
T016206, T024094 and T026435.

\newpage

\begin{table}                                                              
\begin{center}                                                
\begin{tabular}{|c|c|c|c|c|c|c|}\hline
    &   &    &  &  &  &                   \\
    $\alpha$  & ${\Delta \eta}$ & $T_0$ (GeV) &  $y_0$ &
  $\langle u_\rt\rangle$ &  $\langle \frac{\Delta T}{T}\rangle $  
   & $\chi^2$/NDF     \\
      &      &     &  &  &  &              \\
\hline
      &      &      &   &  &   &         \\
  0.26 & 1.36  &  0.140 & 0.082  &  0.20 & 0.71 &  \\
   $\pm$ & $\pm$ & $\pm$ & $\pm$ & $\pm$ & $\pm$ & 642/683 \\
 0.02 & 0.02 & 0.003 & 0.006 & 0.07 & 0.14 &        \\
    &    &     &     &  &  &         \\
\hline
\end{tabular}
\end{center}
\caption{ 
Fit results according to parametrization (1) for $|y|<1.5$.
}       
\end{table}

\vfill

~~~

\newpage

\noindent
{\bf Figure Captions}
\vs 5mm
\begin{itemize}
\item[Fig. 1]
The rapidity distributions of centrally produced pions ($|y|<1.5$) 
for different $m_\rt$-slices given. The curves
are the fit results {obtained} according to parameterization (6).

\item[Fig. 2]
The $(1/m_\rt)$-dependence of ($\Delta y)^2$ for inclusive
$\pi^-$ meson rapidity distributions at $|y|<1.5$. The straight line is 
the fit result according to parametrization (2).
\item[Fig. 3]
The $m_\rt$-distribution of centrally produced pions for different $y$-slices,
as indicated. 
The solid lines  are the fit results {obtained}
according to parameterization (7).
\item[Fig. 4]
$T_{\eff}$ as a function of $y$ fitted according to parametrization (8).
\item[Fig. 5]
The reconstructed $S(t,z)$ emission function in arbitrary units, as a
function of time $t$ and longitudinal coordinate $z$. 
The best fit parameters 
of $\Delta \eta = 1.36$, $y_0 = 0.082$, $\Delta\tau = 1.3$ fm/$c$ and 
$\tau_\rf = 1.4$ fm/$c$ are used to obtain this plot.
\end{itemize}

\ej

\begin{figure}[th]
\epsfig{figure=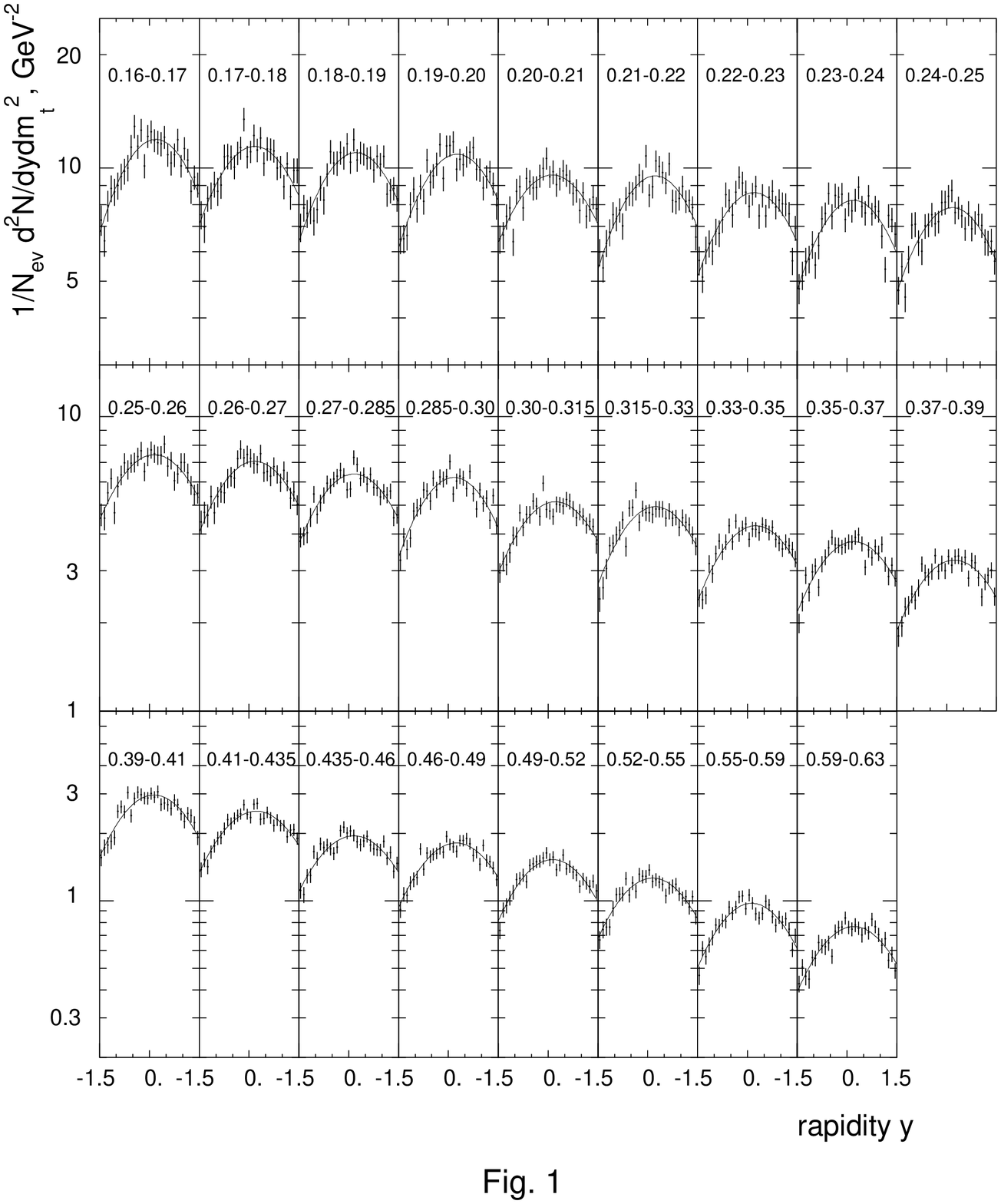,width=15.0cm}
\end{figure}

\ej

\begin{figure}[th]
\noindent 
\centering\epsfig{figure=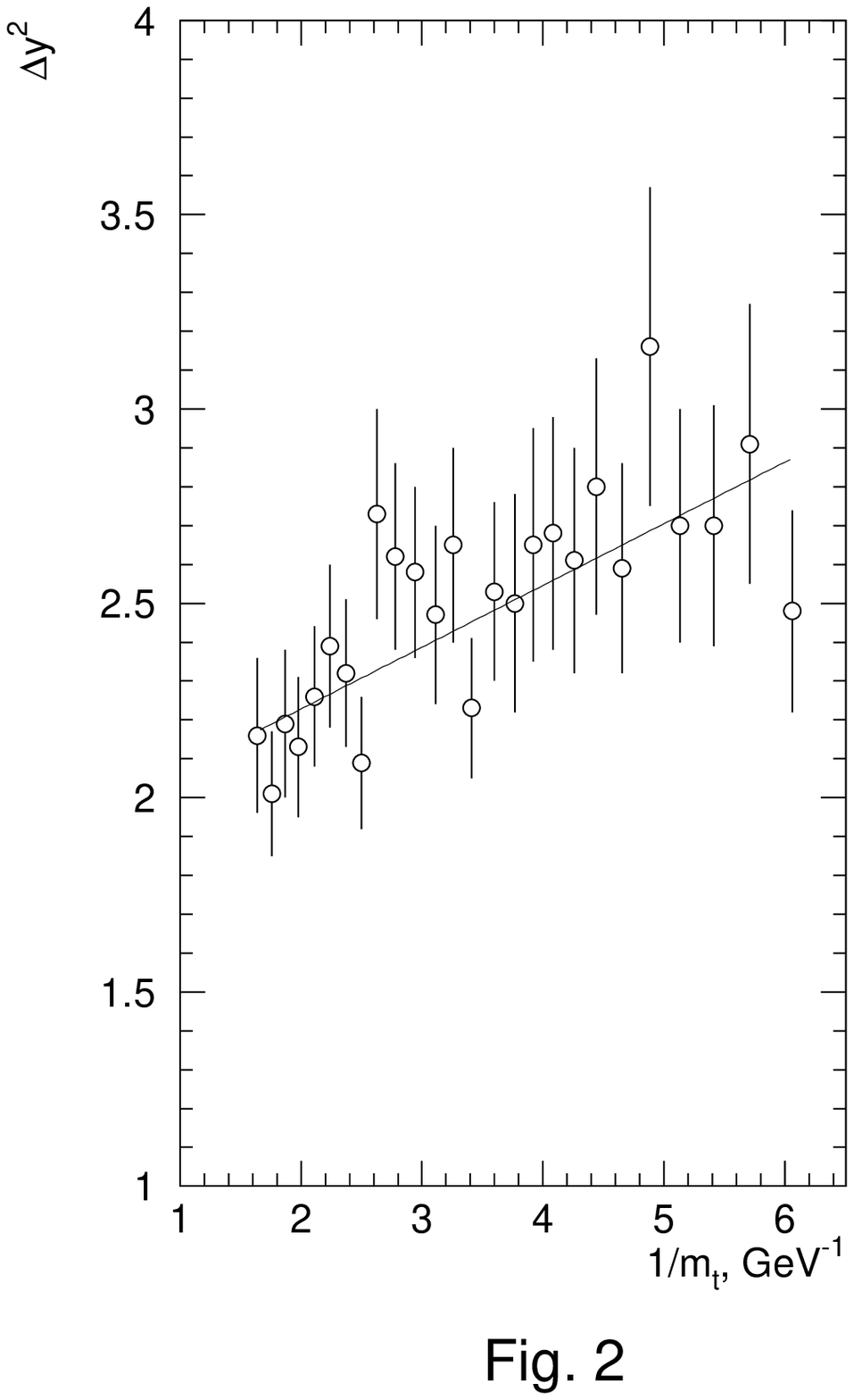}%,width=8cm}
\end{figure}

\ej

\begin{figure}[th]
\epsfig{figure=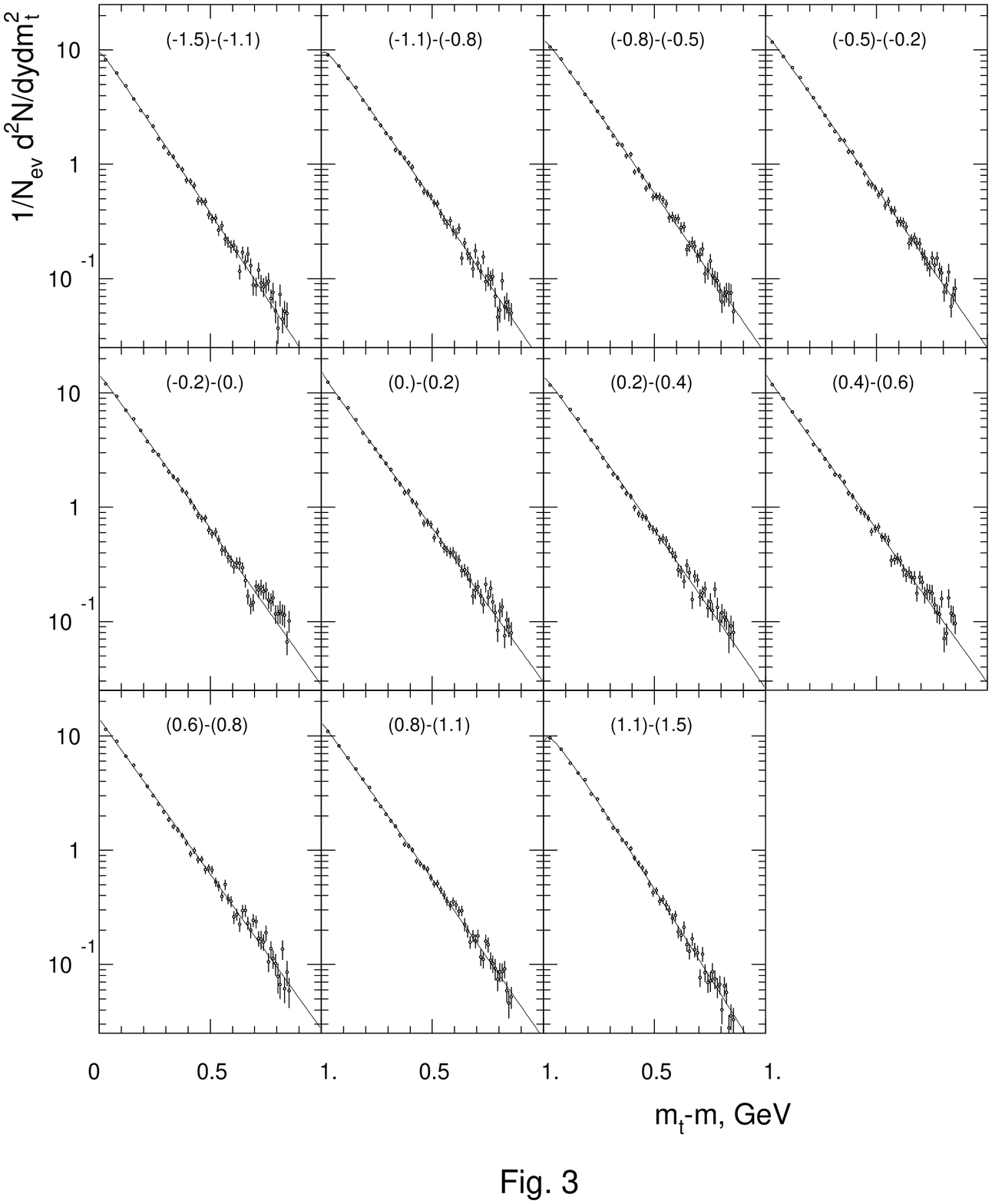,width=15.0cm}
\end{figure}

\ej

\begin{figure}[th]
    \centering\epsfig{figure=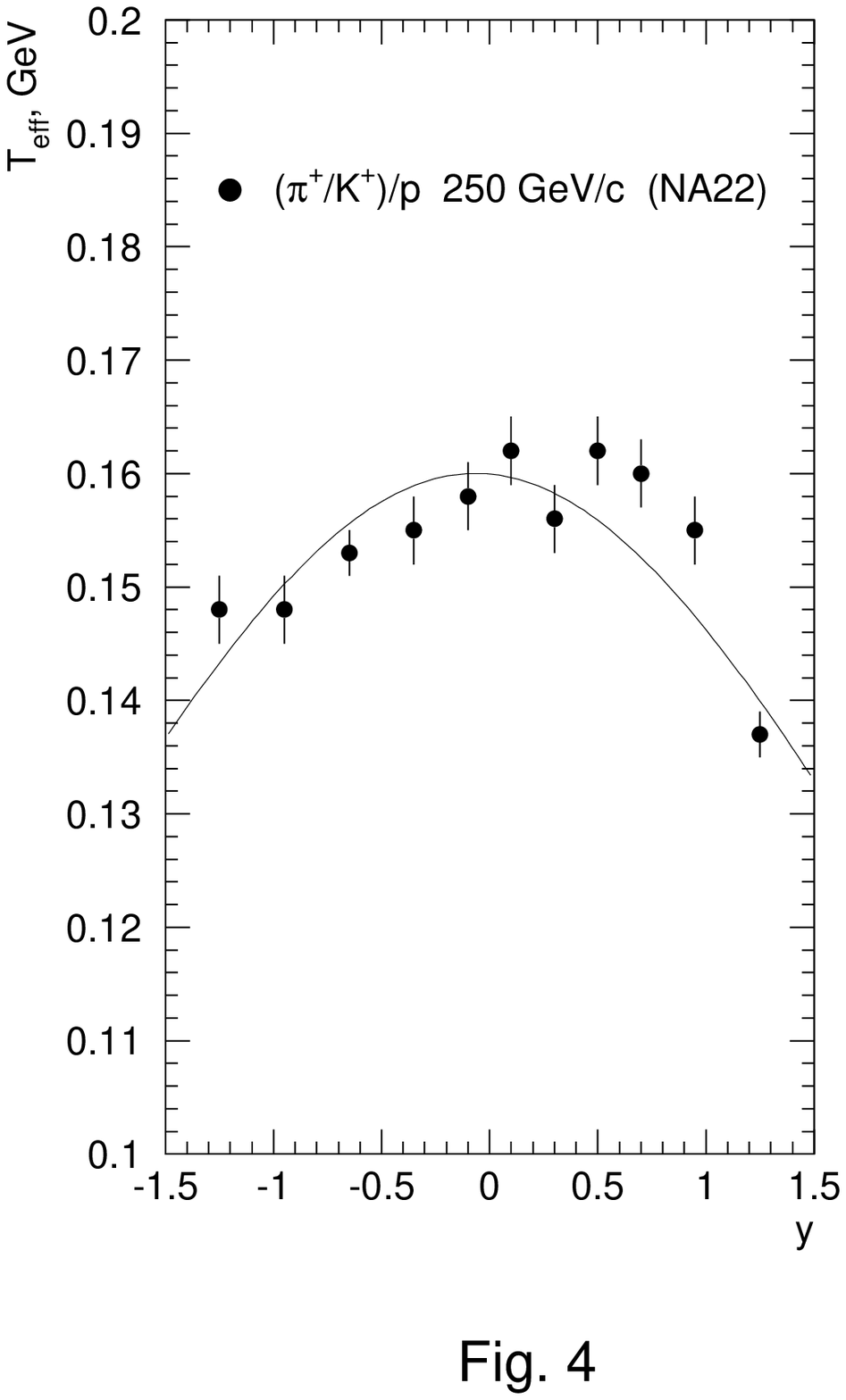}%,width=6.0cm}
\end{figure}

\ej

\begin{figure}[th]
    \centering\epsfig{figure=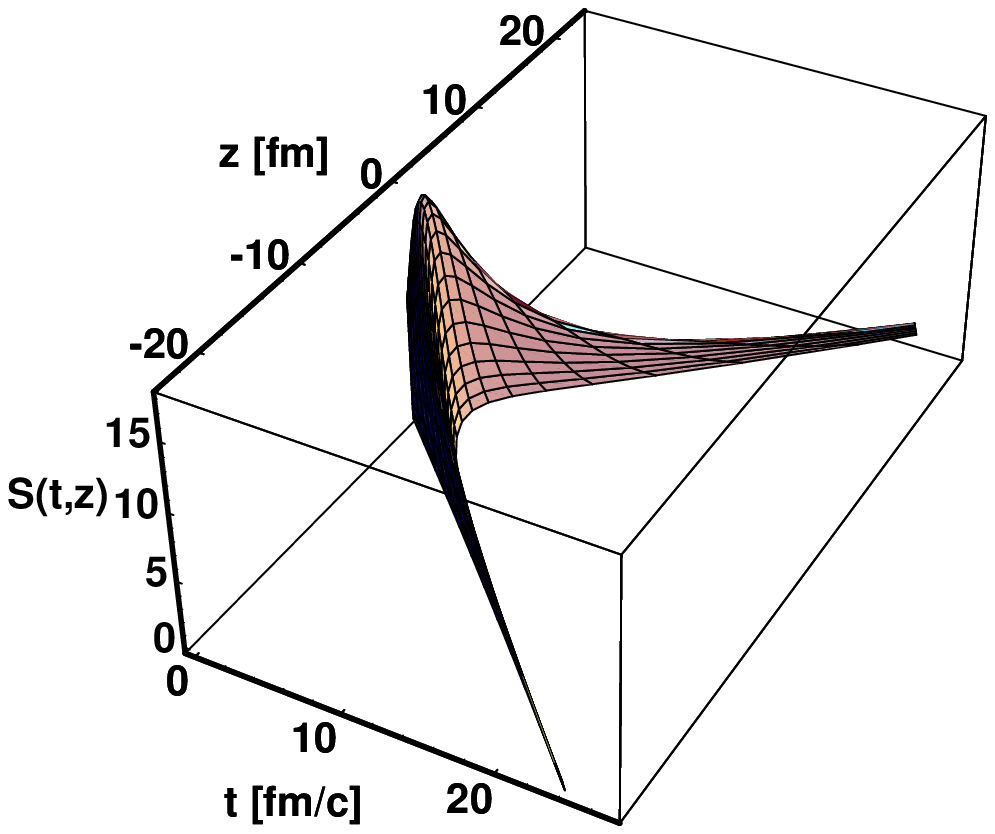}%,width=6.0cm}
\vs 8mm
{\sf \huge Fig.~5}
\end{figure}

\end{document}

%%%%%%%%%%%%%%%%%%%%%%
% End of sprocl.tex  %
%%%%%%%%%%%%%%%%%%%%%%